\documentclass[cameraready]{IEEEtran}
\IEEEoverridecommandlockouts
\usepackage{cite}
\usepackage{amsmath,amssymb,amsfonts}
\usepackage{algorithmic}
\usepackage{graphicx}
\usepackage{textcomp}
\usepackage{xcolor}
\usepackage{booktabs}
\usepackage{multirow}
\usepackage{url}
\usepackage{hyperref}
\usepackage{diagbox}

\hypersetup{
    pdfauthor={Anonymous Authors},
    pdftitle={Making Separation-First Multi-Stream Audio Watermarking Feasible via Joint Training},
    pdfsubject={Double-blind Submission},
    pdfkeywords={audio watermarking, source separation, multi-stream, joint training},
    pdfproducer={LaTeX},
    pdfcreator={pdfTeX}
}

\def\BibTeX{{\rm B\kern-.05em{\sc i\kern-.025em b}\kern-.08em
    T\kern-.1667em\lower.7ex\hbox{E}\kern-.125emX}}

\begin{document}

\title{Making Separation-First Multi-Stream Audio Watermarking Feasible via Joint Training}

\author{
\IEEEauthorblockN{
Houmin Sun\IEEEauthorrefmark{1}\textsuperscript{*},
Zi Hu\IEEEauthorrefmark{1}\textsuperscript{*},
Linxi Li\IEEEauthorrefmark{3},
Yechen Wang\IEEEauthorrefmark{3},
Liwei Jin\IEEEauthorrefmark{3},
Carsten Maple\IEEEauthorrefmark{4}\textsuperscript{\textdagger},
Ming Li\IEEEauthorrefmark{2}\textsuperscript{\textdagger}
}

\IEEEauthorblockA{
\IEEEauthorrefmark{1}Digital Innovation Research Center,
Duke Kunshan University, Kunshan, China\\
\IEEEauthorrefmark{2}School of Artificial Intelligence,
The Chinese University of Hong Kong, Shenzhen, China\\
\IEEEauthorrefmark{3}OfSpectrum, Inc., Los Angeles, USA\\
\IEEEauthorrefmark{4}University of Warwick, Coventry, United Kingdom\\
Email: ming.li.cuhksz@gmail.com
}

\thanks{\textsuperscript{*}Equal contribution.}
\thanks{\textsuperscript{\textdagger}Corresponding author: Ming Li, Carsten Maple.}
}

\maketitle
\begin{abstract}
Modern audio is created by mixing stems from different sources, raising the question: can we independently watermark each stem and recover all watermarks after separation? We study a separation-first, multi-stream watermarking framework --embedding distinct information into stems using unique keys but a shared structure, mixing, separating, and decoding from each output. A naive pipeline (robust watermarking + off-the-shelf separation) yields poor bit recovery, showing robustness to generic distortions does not ensure robustness to separation artifacts. To enable this, we study separation-aware watermarking in a controlled verification pipeline, where the separator is part of the detector and can be selected or optimized together with the watermarking system. Experiments on speech+music and vocal+accompaniment mixtures show substantial gains in post-separation recovery while maintaining perceptual quality.  \footnote{Demo samples are provided at \url{https://anonymous114810.github.io/SF-MSAW-JT/}}.

\end{abstract}


% ================== Introduction ==================
\section{Introduction}

Modern audio production increasingly relies on generative models, neural editing tools, and stem-based workflows. A released waveform may therefore combine speech, vocals, accompaniment, or sound effects originating from different models, creators, or rights holders. This raises a practical provenance question: can each source stream carry its own watermark, and can these watermarks still be recovered after the streams are mixed and later separated? 

Traditional audio watermarking \cite{Bender1996Techniques} method are usually fragile. Existing neural audio watermarking systems are typically designed for single-carrier decoding, where the detector receives the watermarked audio after conventional channel distortions such as noise, filtering, compression, or resampling \cite{liu2023dear, Singh24_SilentCipher}. This setting does not directly address multi-stream mixtures. If several independently watermarked stems are mixed into one waveform, a verifier must first recover source-specific estimates and then decode the corresponding payload from each separated stem.

A natural solution is to apply a music source separator, such as Demucs~\cite{demucs}, TF-GridNet~\cite{Wang23_TFGridNet}, Open-Unmix~\cite{openunmix}, or Spleeter~\cite{spleeter}, before watermark decoding. However, source separation is not a lossless transformation. Even when the separated stems are perceptually acceptable, the separator may introduce source leakage, phase inconsistency, spectral smearing, and other structured artifacts that corrupt watermark-bearing cues~\cite{demucsartificats}. As a result, robustness to generic acoustic distortions does not necessarily imply robustness to separation-induced distortions.

In this work, we study separation-first multi-stream audio watermarking as a controlled known-key payload recovery problem. The separator is treated as part of the verification pipeline rather than as an arbitrary unknown pre-processing step, allowing the watermarking system and separator to be selected or optimized together at detection time. We evaluated whether separation-aware training can improve post-separation payload recovery while preserving perceptual quality and separation integrity.

Our contributions are summarized as follows:
\begin{itemize}
    \item We formulate separation-first multi-stream audio watermarking as a provenance-oriented known-key payload recovery task, where multiple stems in the same mixture carry independent payloads and are decoded after source separation.
    \item We show that strong neural watermarking baselines degrade severely after source separation, indicating that generic robustness does not translate directly to separation robustness.
    \item We introduce separation-aware training strategies that expose the watermarking system to the separation channel and substantially reduce post-separation BER while maintaining perceptual quality.
    \item We analyze separator--watermark compatibility, showing that post-separation watermark recovery depends on the match between the separator architecture and the watermarking model, rather than on separator quality alone.
\end{itemize}

\section{Related Work}

\subsection{Neural Audio Watermarking}

Audio watermarking embeds recoverable information into audio while preserving perceptual quality. Classical methods use signal-processing operations such as phase coding, spread-spectrum embedding, and low-bit coding~\cite{Bender1996Techniques}. Recent neural methods, including WavMark~\cite{Chen23_WavMark}, MaskMark~\cite{OReilly24_MaskMark}, SilentCipher~\cite{Singh24_SilentCipher}, and AURA~\cite{aura2026}, improve payload capacity and robustness against common distortions such as noise, filtering, compression, and resampling. These systems are mainly designed for single-carrier or mixture-level decoding. In contrast, our work considers a separation-first setting, where independently watermarked stems are mixed and then separated before per-stem payload recovery.

\subsection{Source Separation and Watermark Preservation}

Modern source separation systems such as Demucs~\cite{demucs}, TF-GridNet~\cite{Wang23_TFGridNet}, Open-Unmix~\cite{openunmix}, and Spleeter~\cite{spleeter} can produce perceptually plausible stems from a mixture. However, separation remains an ill-posed reconstruction problem and may introduce source leakage, phase inconsistency, spectral smearing, and temporal artifacts~\cite{demucsartificats}. These artifacts are particularly harmful to watermark decoding because watermark cues may depend on fine-grained signal structures that are not preserved by separators optimized only for audio reconstruction quality. 
\subsection{Source Mixing and Separation Robust Steganography}

The closest prior work is the source mixing and separation robust audio steganography method of Takahashi et al.~\cite{takahashi}. They embed secret messages into individual musical sources, mix them with other sources, apply a pre-trained Demucs separator, and recover the messages from the separated outputs. They also show that multiple sources can carry independent messages, using a time-domain concealer-decoder architecture trained with curriculum learning.

Following the distinction noted by Takahashi et al.~\cite{takahashi}, steganography emphasizes covert communication and imperceptibility, whereas watermarking emphasizes robustness for ownership protection and verification. Our work also differs in focus and evaluation. We study provenance-oriented audio watermarking under known-key payload recovery rather than general covert communication. Instead of designing a new source-specific steganographic concealer-decoder, we investigate whether modern neural watermarking systems can support a separation-first verification pipeline, using AURA as a key-conditioned backbone with distinct keys and payloads for different stems. We further compare against recent watermarking baselines, including WavMark, AudioSeal, and AURA, evaluate speech+music and vocal+accompaniment mixtures under additional acoustic attacks, and analyze perceptual quality, separation integrity, and separator-family compatibility with Demucs \cite{demucs}, HTDemucs \cite{Rouard23_Demucs}, and TF-GridNet \cite{Wang23_TFGridNet}.

% ================== Method ==================
\section{Method}

\subsection{Problem Setup: Separation-First Multi-Stream Watermarking}

We consider mixtures formed by multiple stems (e.g., vocal and accompaniment) that may originate from different sources.
Each stem carries an independent watermark embedded with a distinct secret key.
Given stems $\{x_i\}$, watermarks are embedded independently to obtain $\{\tilde{x}_i\}$.
The watermarked stems are mixed into a single waveform:
\begin{equation}
y = \sum_i \tilde{x}_i .
\end{equation}
A source separator $S(\cdot)$ is applied to the mixture to obtain separated estimates $\{\hat{x}_i\} = S(y)$.
The corresponding watermark is then decoded independently from each separated stem $\hat{x}_i$.

\subsection{Watermarking Backbone}

We adopt AURA~\cite{aura2026} as our watermarking backbone, which utilizes a FiLM-conditioned \cite{FiLM} Conformer-based~\cite{conformer} encoder for watermark embedding and a RobustDNN-style \cite{PAVLOVIC2022103381} convolutional decoder for bit extraction. We also tested other neural watermarking frameworks, including AudioSeal~\cite{audioseal} and WavMark~\cite{Chen23_WavMark}, but only selected AURA for joint training because it provided the best perceptual quality among the candidates under our preliminary separation-first evaluation.

\begin{figure*}[t]
    \centering
    \vspace{-0.5cm} % 【榨干1】强行把图片往页面最顶部的边缘推，吃掉顶部留白
    
    % 把宽度稍微收一点点，高度自然就降下来了
    \includegraphics[width=0.7\textwidth,height=0.4\linewidth,keepaspectratio=false]{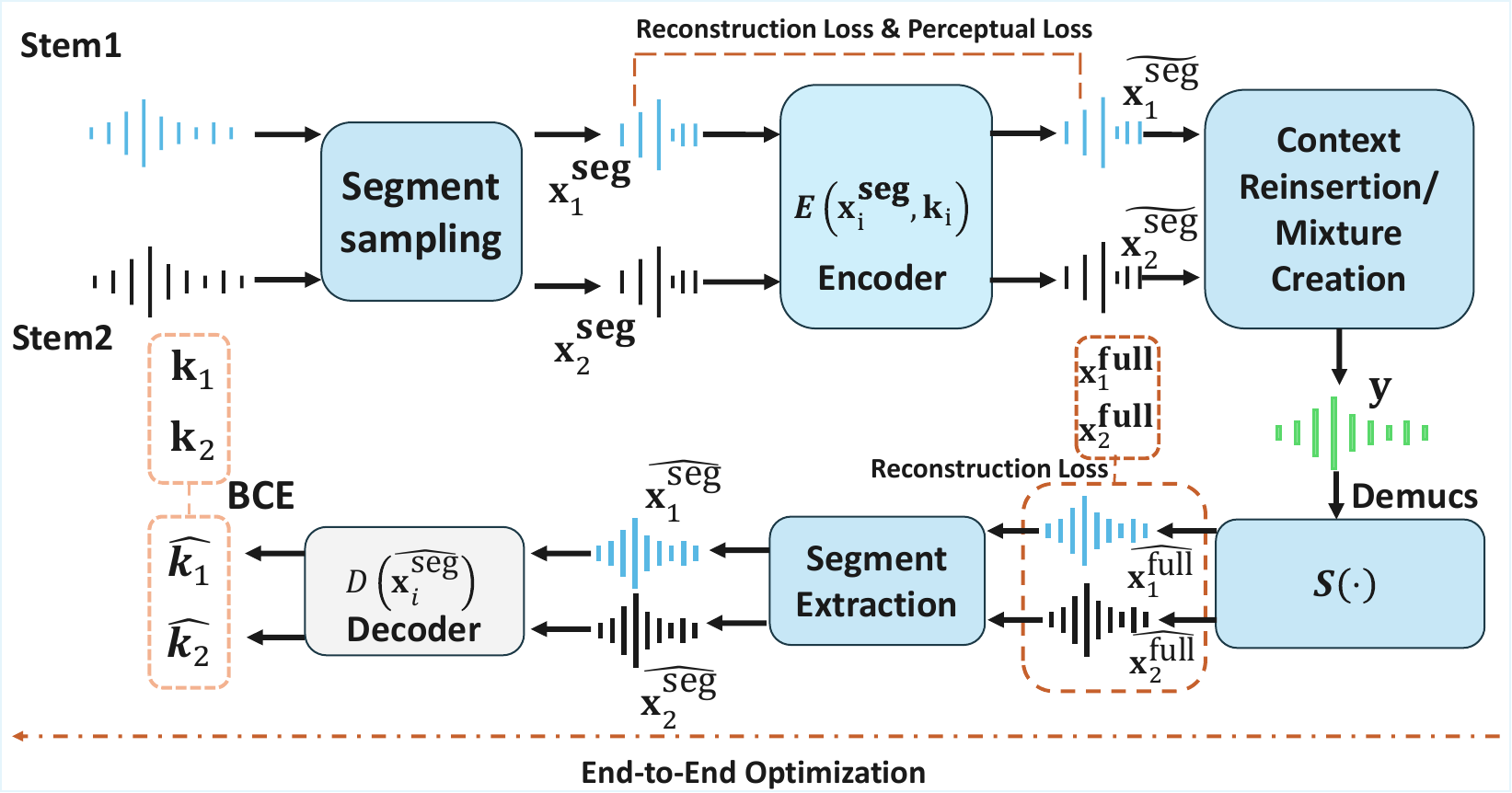} 
    
    %\vspace{-0.3cm} % 【榨干2】让 Caption 紧紧贴着图的下边缘
    \caption{The joint training pipeline for the proposed separation-first audio watermarking framework. In our experiments, \({\mathbf{i}} = 1,2\)}
    \label{fig:pipeline}
    
    %\vspace{-0.6cm} % 【榨干3】核心！强行把下方的正文文字往上提，吃掉海量版面
\end{figure*}

% \subsubsection{Encoder.}

% Given an $n$-second audio segment and a binary secret key, the encoder produces a bounded watermark perturbation in the time--frequency domain.
% The audio is transformed into a magnitude spectrogram, while the key is embedded into a conditioning vector.
% A stack of Conformer~\cite{conformer} blocks, equipped with Feature-wise Linear Modulation (FiLM) layers \cite{FiLM} immediately after the LayerNorm of each of its four core modules (two FFNs, MHSA, and CONV), predicts a real-valued modification mask applied to the magnitude spectrogram. The modified spectrogram is then inverted to the time domain using the original phase to produce the watermarked waveform.

% \subsubsection{Decoder.}
% The decoder takes a potentially distorted audio segment and predicts the embedded bits.
% It follows a RobustDNN-style~\cite{PAVLOVIC2022103381} architecture operating directly on the waveform, using multiple 1D convolutional layers to extract watermark-related features before bit prediction.

\subsection{Source Separation Network}

We use Demucs~\cite{demucs} as the primary source separation backbone in our framework. Demucs is a time-domain separator and is therefore well aligned with our separation-first watermarking setting, where preserving fine temporal structures is important for post-separation payload recovery. We adopt the standard Demucs architecture without modification, so that the observed improvements can be attributed to the proposed separation-aware training pipeline rather than architectural changes to the separator, as detailed in Section \ref{pipeline}. To examine whether the proposed pipeline generalizes beyond this primary backbone, we further conduct extension experiments with HTDemucs and TF-GridNet, as detailed in Section~\ref{Generalization}.

\subsection{Impact of Separation on Watermarking}
\label{impact of separation}
Source separation is inherently lossy~\cite{lossymusic}. Audio separators often introduce artifacts such as phase misalignment, spectral smearing~\cite{vincent2006performance}, and cross-source interference (e.g., residual vocals in the separated accompaniment track)-limitations explicitly acknowledged by the authors of Demucs themselves~\cite{demucsartificats}. These artifacts act as non-stationary, structured noise that is fundamentally different from the generic distortions (e.g., noise addition, compression) against which most watermarking systems are robust. Consequently, a watermark decoder trained only for generic robustness fails to decode correctly from separated stems, as shown in our experiments (Section \ref{exp}). This mismatch necessitates a co-design approach where the watermarking system learns to be robust to the specific distortion profile introduced by the separator, and the separator is encouraged to preserve features critical for watermark recovery.

\subsection{Joint Training Pipeline}
\label{pipeline}
To bridge the robustness gap identified in Section \ref{impact of separation}, we propose an end-to-end differentiable training framework that jointly trains the watermarking encoder/decoder and the Demucs separator through an alternating optimization strategy. The separator is treated as a differentiable separation channel during watermark learning, allowing the watermark decoding loss to be back-propagated through its operations. However, the separator parameters are kept fixed in this update; the BCE decoding loss updates only the watermark encoder and decoder. This encourages the watermarking model to learn embeddings that remain decodable after the specific non-linear distortions introduced by source separation. Each training iteration contains two sequential sub-updates. First, in the watermark update, the separator is used as a differentiable but parameter-frozen proxy channel, and the decoding loss is back-propagated through the separated outputs to update the watermark encoder and decoder. Second, in the separator update, the watermark encoder and decoder are frozen, and the separator is updated solely using the waveform-domain separation loss. Thus, both modules are optimized within each iteration, but the separator shapes the watermark gradients only through the computation graph and is not directly updated by the BCE decoding loss.

The pipeline for each training iteration is detailed below and illustrated in Figure~\ref{fig:pipeline}:
\begin{enumerate}
    \item \textbf{Segment Sampling and Watermark Embedding.} For each stem (e.g., vocal and accompaniment), we sample a full-length audio clip (approximately 20 seconds). From each clip, a \textbf{2-second segment} is cropped to serve as the carrier for watermarking. The 2-second duration represents a standard trade-off in audio watermarking, providing sufficient temporal context (\(L = 88200\) samples at \(44.1\) kHz) for encoding a payload of \(32\) bits while maintaining computational efficiency for \textit{AURA*}. Using our key-conditioned encoder \(E(\cdot, \mathbf{k}_i)\), we embed a unique binary watermark \(\mathbf{b}_i\) into each stem's segment, conditioned on a distinct, randomly generated secret key \(\mathbf{k}_i\). This yields watermarked segments \(\tilde{\mathbf{x}}_i^{\text{seg}} = E(\mathbf{x}_i^{\text{seg}}, \mathbf{k}_i)\).

    \item \textbf{Context Reinsertion.} The watermarked 2-second segments \(\tilde{\mathbf{x}}_i^{\text{seg}}\) are spliced back into their original temporal positions within the 20-second, unwatermarked stems \(\mathbf{x}_i^{\text{full}}\). This produces full-length watermarked stems \(\tilde{\mathbf{x}}_i^{\text{full}}\). This step is critical because Demucs is trained on and expects full-track mixtures. Processing an isolated, short segment may not engage the separator's long-range modeling capabilities (e.g., the bidirectional LSTMs) in the same way as processing a segment within its natural auditory context, which includes preceding and subsequent harmonics, transients, and temporal evolution. Without reinsertion, the separator receives an unnaturally short clip whose boundary conditions and temporal statistics differ from full-track music separation. Reinsertion preserves the local watermark segment while maintaining realistic long-context separation behavior.

    \item \textbf{Mixture Creation.} The full-length watermarked stems \(\tilde{\mathbf{x}}_i^{\text{full}}\) are summed to create the mixture \(\mathbf{y} = \sum_i \tilde{\mathbf{x}}_i^{\text{full}}\). Prior to summation, we \textbf{apply loudness normalization to each stem individually, targeting -16 LUFS (Loudness Units Full Scale)} using the \texttt{pyloudnorm} library. This simulates standard audio mixing practice, ensuring a balanced mixture where no single source dominates. A balanced mixture simplifies the initial separation task, leading to cleaner estimates and reducing a source of interference for the watermark decoder.

    \item \textbf{Separation and Decoding.} We distinguish the secret conditioning key \(\mathbf{k}_i\), which controls the embedding pattern, from the payload bits \(\mathbf{b}_i\), which represent the recoverable watermark message. The decoder predicts \(\hat{\mathbf{b}}_i\) from each separated stem. The mixture \(\mathbf{y}\) is fed into the Demucs separator \(S(\cdot)\), which produces estimates for each stem: \(\hat{\mathbf{x}}_i^{\text{full}} = S_i(\mathbf{y})\). From these separated full-length tracks, we extract the 2-second slices \(\hat{\mathbf{x}}_i^{\text{seg}}\) that are temporally aligned with the originally watermarked segments. Finally, each \(\hat{\mathbf{x}}_i^{\text{seg}}\) is passed to the watermark decoder \(D(\cdot)\) to recover the predicted bit sequence \(\hat{\mathbf{b}}_i = D(\hat{\mathbf{x}}_i^{\text{seg}})\).

%     \item \textbf{Alternating update} During the watermark update, we freeze \(S\) and update \(E,D\) with
% \[
% \min_{\theta_E,\theta_D} \mathcal{L}_{\mathrm{wm}} =
% \mathcal{L}_{\mathrm{bce}} + \lambda_{\mathrm{stft}}\mathcal{L}_{\mathrm{stft}}
% + \lambda_{\mathrm{spec}}\mathcal{L}_{\mathrm{spec}} + \mathcal{L}_{\mathrm{perc}} .
% \]
% During the separator update, we freeze \(E,D\) and update \(S\) with
% \[
% \min_{\theta_S} \mathcal{L}_{\mathrm{sep}} .
% \]
    \item \textbf{Alternating update.} Finally, the task-specific losses (detailed in Section \ref{lfunc}) are calculated. Applying the alternating strategy described above, we first back-propagate the watermark loss \(\mathcal{L}_{\mathrm{wm}}\) through the entire differentiable pipeline to update \(E\) and \(D\), and subsequently use the separation loss \(\mathcal{L}_{\mathrm{sep}}\) to update \(S\).

\end{enumerate}
The entire pipeline is fully differentiable, enabling end-to-end gradient propagation from the decoding loss at the final step back through the separator and to the watermark encoder. 

To isolate the source of the training gain, we further conducted module-freezing ablations in which only selected components were allowed to update. In particular, when the separator was kept fixed, BER still decreased substantially as long as the BCE decoding loss was back-propagated through the separator outputs to update the watermark encoder and decoder. This indicates that the critical ingredient is allowing the watermarking loss to pass through the separator-induced distortion path, rather than directly updating the separator with the BCE loss.

Nevertheless, we adopt the full joint-training pipeline as our primary framework because incorporating the separator into training allows it to adapt to watermarked audio and better preserve separation quality for watermark-bearing stems.

\subsection{Datasets and Training Setup}
\label{datasets}
The watermarking model is pre-trained on approximately 150 hours of instrumental music from Free Music Archive (FMA)~\cite{FMA} and 300 hours of speech from the Emilia~\cite{Emilia} dataset. This pre-training phase is conducted on four NVIDIA RTX 4090D GPUs for 48 hours. Additional training uses MUSDB18~\cite{musdb18} dataset. The end-to-end joint training is performed on four NVIDIA L20 GPUs for 72 hours. All audio is resampled to 44.1 kHz, and two-second segments are used as watermark carriers. We use the Adam optimizer with a learning rate of $10^{-4}$ and initialize from pre-trained checkpoints.
\subsection{Loss Functions}
\label{lfunc}
Training is driven by a multi-task objective for both the Demucs~\cite{demucs} and \textit{AURA*}~\cite{aura2026}:
\begin{equation}
    \mathcal{L}_{\mathrm{total}} = 100 \mathcal{L}_{\mathrm{bce}} + \lambda_{\mathrm{stft}} \mathcal{L}_{\mathrm{stft}} + \lambda_{\mathrm{spec}} \mathcal{L}_{\mathrm{spec}} + \mathcal{L}_{\mathrm{perc}}
\end{equation}
Here, $\mathcal{L}_{\mathrm{bce}} = \mathrm{BCE}(\mathbf{b}, \hat{\mathbf{b}})$ ensures reliable bit recovery. $\mathcal{L}_{\mathrm{stft}}$ is a multi-resolution log-magnitude STFT loss across $K\!=\!3$ resolutions (fft sizes: 1024, 2048, 512): $\mathcal{L}_{\mathrm{stft}} = \frac{1}{K} \sum_{k=1}^K | \log(|X_k|\!) - \log(|\tilde{X}_k|\!) |_1$, where $X_k$ and $\tilde{X}_k$ are the original and watermarked spectrograms. $\mathcal{L}_{\mathrm{spec}}$ computes the multi-scale Mel-spectrogram $L_1$ distance. Pre-training follows a two-phase schedule: we initially prioritize bit recovery with minimal spectral weighting and no $\mathcal{L}_{\mathrm{perc}}$. After 20k steps, spectral weights are increased significantly and $\mathcal{L}_{\mathrm{perc}}$ is introduced, maintaining this balance throughout joint training. For imperceptibility, $\mathcal{L}_{\mathrm{perc}}$ combines a BigVGAN discriminator loss \cite{BigVGAN} and a Noise-to-Mask Ratio loss \cite{NMR}: $\mathcal{L}_{\mathrm{nmr}} = \frac{1}{CT} \sum_{c,t} (N_{c,t} / M_{c,t})$, where $N_{c,t}$ and $M_{c,t}$ are the noise and masking patterns across $C$ critical bands and $T$ frames, explicitly penalizing distortions that exceed local time--frequency auditory masking thresholds. Crucially, to preserve separation integrity without collapsing the learned watermarking space, the separator is updated \emph{exclusively} via an independent waveform-domain $L_1$ loss:
\begin{equation}
    \mathcal{L}_{\mathrm{sep}} = | \hat{v} - \mathrm{sg}(\tilde{v}) |_1 + | \hat{a} - \mathrm{sg}(\tilde{a}) |_1
\end{equation}
where $\hat{v}, \hat{a}$ are the separated vocal and accompaniment estimates, and $\mathrm{sg}(\cdot)$ is a stop-gradient operation (\texttt{detach}) on the watermarked targets.

\subsubsection{Attack Simulations}
After 20,000 training steps, we introduce an attack simulator applying 18 random distortions to the mixed, watermarked audio. Attacks are applied to the full mixed waveform before separation, rather than to isolated stems, matching the scenario where a released mixture is transformed before provenance verification. For a structured evaluation, the subsequent robustness results will be presented in Table I according to these three distinct categories: 
\begin{itemize}
    \item \textbf{Basic/Noise:} 
    \begin{itemize}
    \item \textbf{N,PK}  Additive white and peak-normalized pink noise  
    \item \textbf{SP}  0.1\% Random sample suppression 
    \item \textbf{AMP, BST, DK}  Random amplitude scaling in $[-1.0, 1.0]$ alongside fixed 20\% boosting/ducking 
    \item \textbf{QT}  Bit-depth quantization 
    \item \textbf{PHS}  FFT-based global phase rotation 
    \end{itemize}
    \item  \textbf{Filter:} 
    \begin{itemize}
    \item \textbf{LP, BF}  Biquad low-pass (3--6 kHz cutoffs) and band-pass filtering 
    \item \textbf{SM}  Moving-average smoothing with a 2--10 sample window
    \item \textbf{SPAUG}  Random time--frequency spectrogram masking
    \end{itemize}
    \item \textbf{Time/Pitch:} 
    \begin{itemize}
    \item \textbf{RS}  Resampling across diverse rates down to 16 kHz 
    \item \textbf{Echo, RV}  Single echo (100 ms delay) and convolutional reverberation 
    \item \textbf{SPD, PCH, SPCH}  Alongside independent and combined pitch-preserving speed and speed-preserving pitch modifications 
    \end{itemize}
\end{itemize}
During joint training, these simulated attacks force the network to explicitly optimize for watermark detectability and separation fidelity under severe, real-world acoustic degradations.
% ================== Experiments ==================
\begin{table*}[t]
\caption{Average Bit Error Rate (BER, \%) under separation-first evaluation across different benchmarks.}
\label{tab:robustness}
\centering
\small
\renewcommand{\arraystretch}{1.1}
\setlength{\tabcolsep}{4.5pt} % 稍微缩小间距以防溢出
\begin{tabular}{l|cccc|cccc}
\toprule
\multirow{2}{*}{\textbf{Model}} & \multicolumn{4}{c|}{\textbf{Speech + Music Scenarios}} & \multicolumn{4}{c}{\textbf{Vocal + Accompaniment Scenarios}} \\
\cmidrule(lr){2-5} \cmidrule(lr){6-9}
& Origin & Basic/Noise & Filter & Time/Pitch & Origin & Basic/Noise & Filter & Time/Pitch \\
\midrule
WavMark & 35.45 & 37.37 & 38.38 & 41.69 & 30.70 & 31.64 & 33.53 & 37.84 \\
AudioSeal & 26.88 & 27.07 & 29.46 & 37.00 & 19.75 & 20.01 & 24.82 & 34.91 \\
\textit{AURA*} & 15.10 & 16.15 & 17.86 & 35.85 & 19.23 & 19.31 & 20.82 & 34.86 \\
\midrule
\textit{AURA*-Demucs$_{joint}$} & \textbf{0.43} & \textbf{1.07} & \textbf{2.81} & \textbf{20.43} & 15.21 & 17.55 & 19.53 & 31.68 \\
\textit{AURA*-Demucs$_{finetuned}$} & 1.15 & 1.82 & 3.47 & 21.26 & \textbf{1.25} & \textbf{1.82} & \textbf{3.65} & \textbf{31.60} \\
\bottomrule
\end{tabular}
\end{table*}

\section{Experiments}
\label{exp}

\textbf{Model Variants.}
We evaluate several watermarking baselines alongside two separation-aware variants. \textit{AURA*} denotes our implementation of the published AURA watermarking backbone~\cite{aura2026}, trained on the datasets described in Section~\ref{datasets}. \textit{AURA*-Demucs$_{joint}$} denotes the full separation-aware model trained on FMA--Emilia speech+music mixtures. \textit{AURA*-Demucs$_{finetuned}$} denotes the dataset-adapted variant further trained on MUSDB18 vocal/accompaniment mixtures, therefore achieves substantially better BER on the vocal+accompaniment benchmark.

We evaluated WavMark, AudioSeal, \textit{AURA*}, and 2 separation-aware variants -- \textit{AURA*-Demucs$_{joint}$}, and \textit{AURA*-Demucs$_{finetuned}$}---based both on robustness and perceptual quality. Robustness was quantified by the average bit error rate (BER) under the evaluation pipeline described in Section \ref{robustnessspe}, which includes watermark encoding, loudness alignment, mixing, Demucs separation, and per-stem decoding.  All baselines are evaluated using their native operating configurations to avoid disadvantaging any method by forced resampling or non-native payload settings. Specifically, WavMark uses its native sampling rate with a 32-bit payload per 1-second window and a randomly sampled key, AudioSeal uses its native sampling rate with a 16-bit payload per 1-second window, and AURA* uses its 44.1 kHz sampling rate with a 32-bit payload per 2-second window and random keys. All methods are then tested under the same separation-first pipeline, with BER computed over each method's decoded payload bits. All methods share the same evaluation settings as the joint training configuration. Test data consist of a 15-hour speech+music set, synthesized from non-overlapping subsets of Emilia~\cite{Emilia} and FMA~\cite{FMA}, and a 10-hour vocal+accompaniment set extracted from MUSDB18~\cite{musdb18}. All test data were strictly excluded from training datasets. 

\subsection{Evaluation Metrics}

We evaluate audio fidelity using Signal-to-Noise Ratio (SNR) and Scale-Invariant Signal-to-Noise Ratio (SI-SNR) to quantify waveform reconstruction (accounting for gain fluctuations), and ViSQOL~\cite{Chinen20_ViSQOL} for human-perceived quality on a 1–5 Mean Opinion Score (MOS) scale. In order to evaluate the robustness of audio watermarking, we report aggregate BER, computed as the total number of incorrectly decoded bits divided by the total number of decoded bits over all test windows and samples. 

% ================== Table 2 ==================
\begin{table}[t]
  \caption{Perceptual quality comparison without separation.}
  \label{tab:perceptual}
  \centering
  \begin{tabular}{lccc}
    \toprule
    \textbf{Model} & \textbf{SNR} & \textbf{ViSQOL} & \textbf{SI-SNR}\\
    \midrule
    \textit{AURA*} & 30.66 & 4.72 & 33.00 \\
    \textit{AURA*-Demucs$_{joint}$} & 28.77  & 4.72 & 32.80 \\
    \textit{AURA*-Demucs$_{finetuned}$} & 18.56  & 4.51 & 26.79 \\
    \bottomrule
  \end{tabular}
\end{table}

% ================== Table 3 ==================
\begin{table}[t]
\caption{Separation Performance of  Demucs}
\label{tab:perceptual_separation}
\centering
\begin{tabular}{lcccc}
\toprule
\multirow{2}{*}{Timeline} & \multicolumn{2}{c}{Vocal} & \multicolumn{2}{c}{Accompaniment} \\
\cmidrule(lr){2-3} \cmidrule(lr){4-5}
 & SDR & SI-SDR & SDR & SI-SDR \\
\midrule
Before & 9.36 & 8.66 & 12.50 & 12.30 \\
After  & 8.89 & 8.10 & 13.19 & 12.98 \\
\bottomrule
\end{tabular}
\end{table}

\subsection{Perceptual Quality and Separation Integrity}

We evaluate two complementary aspects of audio fidelity: the imperceptibility of embedded watermarks and the integrity of separated stems after joint training.

\textbf{Watermark imperceptibility.} Table~\ref{tab:perceptual} presents the perceptual quality of watermarked audio (without separation). The Joint and Finetuned models achieve SNRs of 28.77 dB and 18.56 dB, respectively, with ViSQOL \cite{Chinen20_ViSQOL} scores above 4.5. As ViSQOL aligns more closely with human auditory perception than waveform reconstruction metrics, these results confirm that the embedded watermarks remain effectively imperceptible. This confirms that our separation-aware training does not compromise the fundamental requirement of audio watermarking: inaudible embedding. The fact that separation-aware models do not outperform single-carrier baselines in this direct encoding/decoding setting is expected, as our joint objective explicitly prioritizes robustness to separation-induced distortions over standalone imperceptibility.

\textbf{Separation integrity.} A critical question is whether optimizing for watermark robustness degrades the core functionality of the Demucs separator. Table~\ref{tab:perceptual_separation} addresses this by comparing separation quality before and after joint training using standard source separation metrics (SDR and SI-SDR). The results demonstrate that separation integrity is well preserved: vocal SDR changes minimally from 9.36 to 8.89, while accompaniment SDR slightly improves from 12.50 to 13.19. We attribute this asymmetry to the interplay between watermark preservation and spectral density. The sparser vocal track forces the separator to retain watermark-bearing perturbations, marginally reducing its SDR. Conversely, the complex accompaniment spectrum allows joint training to effectively fine-tune the separator's instrumental representations, slightly improving its SDR. These negligible differences confirm that our framework enhances watermark robustness without compromising the separator's ability to produce high-quality stems. 

\subsection{Robustness Against Separation}
\label{robustnessspe}
To evaluate whether watermarking methods can preserve decoding accuracy after source separation, we adopt a separation-first evaluation standard aligned with the training pipeline in Section \ref{pipeline}. Watermarks are embedded independently into two-second segments from distinct constituent sources, which are subsequently mixed and separated via Demucs. The Bit Error Rate (BER) is computed using the temporally aligned segments extracted from each separated stem. 

As demonstrated in Table 1, conventional neural watermarking baselines, such as WavMark and AudioSeal, exhibit prohibitively high bit error rates post-separation, with the BER deteriorating to 20--30\% across both benchmarks. The vanilla \textit{AURA*} provides only marginal improvements, implying that robustness to generic acoustic distortions does not inherently translate to resilience against source separation artifacts.

In contrast, the separation-aware variants substantially improve post-separation decoding compared with the generic watermarking baselines, but their behavior differs across domains. On the speech+music benchmark, both variants achieve low BERs after separation, reducing the origin-condition BER to 0.43\% for \textit{AURA*-Demucs$_{joint}$} and 1.15\% for \textit{AURA*-Demucs$_{finetuned}$}. Under additional basic/noise and filtering attacks, both variants remain below 4\% BER, whereas the non-separation-aware baselines remain above 15\%. This confirms that exposing the watermarking system to the separation channel is crucial for robust post-separation decoding.

On the vocal+accompaniment benchmark, however, the two variants exhibit different behavior. The fine-tuned model achieves the strongest performance, reducing the origin-condition BER to 1.25\% and maintaining low BER under basic/noise and filtering attacks. In contrast, the fully joint variant remains less effective in this domain, with a 15.21\% BER in the origin condition. This suggests that unrestricted co-adaptation between the watermarking modules and the separator may not always transfer cleanly across source domains, especially for real musical stems with dense harmonic structure and strong source overlap. The fine-tuned setting, which adapts the separator while keeping the watermarking model fixed, appears to better preserve the learned watermark representation in this case.

Overall, these results show that separation-aware training is necessary for reliable multi-stream watermark recovery after source separation, while also indicating that the choice of adaptation strategy is domain-dependent. Cross-stream interference can be substantially mitigated, but achieving uniformly low BER across different mixture types remains an important design consideration. We also observe that the Time/Pitch category remains the most difficult setting. These transformations alter temporal alignment and phase structure, which can simultaneously affect source separation and bit-level watermark decoding. Therefore, although separation-aware training greatly improves robustness under various of conditions, robustness to time-scale, pitch transformations remains an open challenge.

\subsubsection{Separator Generalization and Domain Compatibility}
\label{Generalization}
A pivotal inquiry is whether a watermarking framework co-optimized with a specific separator (e.g., Demucs) can generalize to alternative separation architectures at inference time. As tabulated in Table~\ref{tab:separator_generalization}, our empirical cross-testing reveals that separation-aware watermarking is highly coupled with the distortion profiles and signal representations of the training channel. Direct model replacement across completely different separator families induces severe decoding degradation.

To trace the boundaries of this compatibility and understand how different separation paradigms affect watermark survival, we extend our joint training pipeline to two distinct architectural alternatives: \textbf{HTDemucs}~\cite{Rouard23_Demucs} (a time-domain, hybrid transformer variation) and \textbf{TF-GridNet} (a state-of-the-art frequency-domain separator).

Our empirical results show a clear contrast between the two separator families. The pipeline consistently converges with \textbf{HTDemucs}, achieving BER and perceptual quality comparable to the default Demucs setup. In contrast, under the same separation-aware training protocol, repeated runs with \textbf{TF-GridNet} did not converge to reliable watermark recovery, with the BER remaining close to chance level (approximately 50\%) after multiple training attempts.

We hypothesize that this discrepancy arises from how the separators reconstruct low-energy time-frequency details. AURA* embeds and decodes payloads through subtle magnitude-spectral patterns, which can be treated by a separator as non-source perturbations rather than source-consistent content. TF-GridNet performs complex spectral mapping and explicitly reconstructs source spectra, so such weak watermark-bearing patterns may be suppressed, redistributed across sources, or decorrelated from the decoder's expected structure. In contrast, Demucs-style and HTDemucs separators combine waveform-domain reconstruction with less direct spectral masking, which may better preserve the residual spectral perturbations learned by AURA* under our training protocol. These results suggest that post-separation watermark recovery depends not only on separation quality, but also on whether the separator's reconstruction bias preserves the specific signal structures used by the watermarking model. 

\begin{table}[t]
\centering
\caption{Separator-family generalization under separation-aware joint training.}
\label{tab:separator_generalization}
\scriptsize
\begin{tabular}{lccc}
\toprule
\textbf{Separator} & \textbf{Domain} & \textbf{BER (\%) $\downarrow$}  \\
\midrule
Demucs & time & 1.10  \\
HTDemucs & time & 1.12 \\
TF-GridNet & complex time-frequency & $\approx 50$  \\
\bottomrule
\end{tabular}
\end{table}

% ================== Conclusion ==================
\section{Conclusion}

We investigated separation-first multi-stream audio watermarking and showed that robustness to common distortions does not necessarily imply robustness to source separation. By incorporating the separator into the training pipeline, our separation-aware models substantially improve post-separation payload recovery while preserving perceptual quality and separation integrity. The results also highlight key limitations: time/pitch transformations remain challenging, and the learned watermarking model does not transfer reliably across arbitrary separator architectures. In particular, compatible time-domain separators such as Demucs and HTDemucs perform better than frequency-domain alternatives in our current pipeline. Future work will explore separator-agnostic watermark cues and extend the framework beyond two-stem mixtures.

\section{Generative AI Use Disclosure}
Large Language Models (LLMs) were used solely for manuscript polishing (e.g., rephrasing and grammar checks) to improve clarity and readability. The LLMs were not used for ideation, methodology, experimental design, data analysis, or result interpretation. All scientific content was produced and verified by the authors.
~\\
~\\

\bibliographystyle{IEEEtran}
\bibliography{mybib}

@article{Bender1996Techniques,
  author  = {Bender, Walter and Gruhl, Daniel and Morimoto, Norishige and Lu, Anthony},
  title   = {Techniques for data hiding},
  journal = {IBM Systems Journal},
  volume  = {35},
  number  = {3.4},
  pages   = {313--336},
  year    = {1996}
}

@article{Chen23_WavMark,
  author  = {Chen, Guangyu and Wu, Yu and Liu, Shujie and Liu, Tao and Du, Xiaoyong and Wei, Furu},
  title   = {WavMark: Watermarking for Audio Generation},
  journal = {arXiv preprint arXiv:2308.12770},
  year    = {2024}
}

@inproceedings{OReilly24_MaskMark,
  author    = {O'Reilly, Patrick and Jin, Zeyu and Su, Jiaqi and Pardo, Bryan},
  title     = {Maskmark: Robust Neuralwatermarking for Real and Synthetic Speech},
  booktitle = {Proc. ICASSP}, 
  year      = {2024},
  pages     = {4650--4654}
}

@inproceedings{Singh24_SilentCipher,
  author    = {Singh, Mayank Kumar and Takahashi, Naoya and Liao, Weihsiang and Mitsufuji, Yuki},
  title     = {SilentCipher: Deep Audio Watermarking},
  booktitle = {Proc. Interspeech},
  year      = {2024},
  pages     = {2235--2239}
}

@article{Rouard23_Demucs,
  author  = {Rouard, Simon and Massa, Francisco and Défossez, Alexandre},
  title   = {Hybrid Transformers for Music Source Separation},
  journal = {arXiv preprint arXiv:2211.08553},
  year    = {2022}
}

@article{Wang23_TFGridNet,
  author  = {Wang, Zhong-Qiu and Cornell, Samuele and Choi, Shukjae and Lee, Younglo and Kim, Byeong-Yeol and Watanabe, Shinji},
  title   = {TF-GridNet: Making Time-Frequency Domain Models Great Again for Monaural Speaker Separation},
  journal = {arXiv preprint arXiv:2209.03952},
  year    = {2023}
}

@article{FiLM,
  author  = {Perez, Ethan and Strub, Florian and de Vries, Harm and Dumoulin, Vincent and Courville, Aaron C.},
  title   = {FiLM: Visual Reasoning with a General Conditioning Layer},
  journal = {arXiv preprint arXiv:1709.07871},
  year    = {2017}
}

@article{PAVLOVIC2022103381,
  author  = {Pavlović, Kosta and Kovačević, Slavko and Djurović, Igor and Wojciechowski, Adam},
  title   = {Robust speech watermarking by a jointly trained embedder and detector using a DNN},
  journal = {Digital Signal Processing},
  volume  = {122},
  pages   = {103381},
  year    = {2022}
}

@article{demucsartificats,
  author  = {Défossez, Alexandre},
  title   = {Hybrid Spectrogram and Waveform Source Separation},
  journal = {arXiv preprint arXiv:2111.03600},
  year    = {2022}
}

@article{lossymusic,
  author  = {Cano, Estefania and FitzGerald, Derry and Liutkus, Antoine and Plumbley, Mark D. and Stöter, Fabian-Robert},
  title   = {Musical Source Separation: An Introduction},
  journal = {IEEE Signal Processing Magazine}, 
  volume  = {36},
  number  = {1},
  pages   = {31--40},
  year    = {2019}
}

@article{FMA,
  author  = {Defferrard, Michaël and Benzi, Kirell and Vandergheynst, Pierre and Bresson, Xavier},
  title   = {FMA: A dataset for music analysis},
  journal = {arXiv preprint arXiv:1612.01840},
  year    = {2016}
}

@article{Emilia,
  author  = {He, Haorui and Shang, Zengqiang and others},
  title   = {Emilia: An Extensive, Multilingual, and Diverse Speech Dataset for Large-Scale Speech Generation},
  journal = {arXiv preprint arXiv:2407.05361},
  year    = {2024}
}

@misc{musdb18,
  author  = {Rafii, Zafar and Liutkus, Antoine and Stöter, Fabian-Robert and Mimilakis, Stylianos Ioannis and Bittner, Rachel},
  title   = {The MUSDB18 corpus for music separation},
  year    = {2017}
}

@article{Chinen20_ViSQOL,
  author  = {Chinen, Michael and Lim, Felicia S. C. and Skoglund, Jan and Gureev, Nikita and O'Gorman, Feargus and Hines, Andrew},
  title   = {ViSQOL v3: An Open Source Production Ready Objective Speech and Audio Metric},
  journal = {arXiv preprint arXiv:2004.09584},
  year    = {2020}
}

@article{BigVGAN,
  author  = {Lee, Sang-gil and Ping, Wei and Ginsburg, Boris and Catanzaro, Bryan and Yoon, Sungroh},
  title   = {BigVGAN: A Universal Neural Vocoder with Large-Scale Training},
  journal = {arXiv preprint arXiv:2206.04658},
  year    = {2023}
}

@inproceedings{NMR,
  author    = {Moritz, Martin and Olán, Toni and Virtanen, Tuomas},
  title     = {Noise-to-Mask Ratio Loss for Deep Neural Network Based Audio Watermarking},
  booktitle = {Proc. IS2},
  year      = {2024},
  pages     = {1--6} 
}

@article{conformer,
  author  = {Gulati, Anmol and Qin, James and Chiu, Chung-Cheng and Parmar, Niki and Zhang, Yu and Yu, Jiahui and Han, Wei and Wang, Shibo and Zhang, Zhengdong and Wu, Yonghui and Pang, Ruoming},
  title   = {Conformer: Convolution-augmented Transformer for Speech Recognition},
  journal = {arXiv preprint arXiv:2005.08100},
  year    = {2020}
}

@article{openunmix,
  author  = {Stöter, Fabian-Robert and Uhlich, Stefan and others},
  title   = {Open-Unmix - A Reference Implementation for Music Source Separation},
  journal = {Journal of Open Source Software},
  volume  = {4},
  number  = {41},
  pages   = {1667},
  year    = {2019}
}

@article{spleeter,
  author  = {Hennequin, Romain and Khlif, Anis and others},
  title   = {Spleeter: a fast and efficient music source separation tool with pre-trained models},
  journal = {Journal of Open Source Software},
  volume  = {5},
  number  = {50},
  pages   = {2154},
  year    = {2020}
}

@article{vincent2006performance,
  author  = {Vincent, Emmanuel and Gribonval, R{\'e}mi and F{\'e}votte, C{\'e}dric},
  title   = {Performance measurement in blind audio source separation},
  journal = {IEEE Trans. Audio, Speech, Lang. Process.},
  volume  = {14},
  number  = {4},
  pages   = {1462--1469},
  year    = {2006}
}

@article{liu2023dear,
  author  = {Liu, Chang and Zhang, Jie and others},
  title   = {DeAR: A Deep-learning-based Audio Re-recording Resilient Watermarking},
  journal = {arXiv preprint arXiv:2212.02339},
  year    = {2023}
}

@inproceedings{takahashi,
  author    = {Naoya Takahashi and Mayank Kumar Singh and Yuki Mitsufuji},
  title     = {Source Mixing and Separation Robust Audio Steganography},
  booktitle = {Proc. Interspeech 2021},
  pages     = {4554--4558},
  year      = {2021}
}

@INPROCEEDINGS{aura2026,
  author={Li, Linxi and Jin, Liwei and Wang, Yechen and Sun, Houmin and Hu, Zi and Maple, Carsten},
  booktitle={ICASSP 2026 - 2026 IEEE International Conference on Acoustics, Speech and Signal Processing (ICASSP)}, 
  title={AURA: A Stegaformer-Based Scalable Deep Audio Watermark with Extreme Robustness}, 
  year={2026},
  volume={},
  number={},
  pages={16522-16526},
  doi={10.1109/ICASSP55912.2026.11462806}}

@misc{audioseal,
      title={Proactive Detection of Voice Cloning with Localized Watermarking}, 
      author={Robin San Roman and Pierre Fernandez and Alexandre Défossez and Teddy Furon and Tuan Tran and Hady Elsahar},
      year={2024},
      eprint={2401.17264},
      archivePrefix={arXiv},
      primaryClass={cs.SD},
      url={https://arxiv.org/abs/2401.17264}, 
}

@misc{demucs,
      title={Demucs: Deep Extractor for Music Sources with extra unlabeled data remixed}, 
      author={Alexandre Défossez and Nicolas Usunier and Léon Bottou and Francis Bach},
      year={2019},
      eprint={1909.01174},
      archivePrefix={arXiv},
      primaryClass={cs.SD},
      url={https://arxiv.org/abs/1909.01174}, 
}
\end{document}